\documentstyle[12pt]{article}
\begin{document}

\title{
\begin{flushright}
{\small SMI-2-93 \\ February, 1993 }
\end{flushright}
\vspace{2cm}
Affine Strings}
\author{I.V. Volovich
\\ Steklov Mathematical Institute,\\ Russian Academy of Sciences,\\
Vavilov st.42, GSP-1,117966, \\ Moscow, Russia }
\date{~}
\maketitle
\begin {abstract}
A new model of bosonic strings  is considered. An action of the
model is the sum of the standard string action and  a term describing
an interaction of a metric with a linear (affine) connection.
The Lagrangian of this interaction is an arbitrary analytic function
$f(R)$ of the scalar curvature. This is a classically integrable model.
The space of classical solutions of the theory consists from
sectors with constant curvature. In each sector the equations
of motion reduce to  the standard string equations  and to an
additional constant curvature equation for the linear connection.
A bifurcation in the space of all Lagrangians takes place.
Quantization of the model is briefly discussed. In a quasiclassical
approximation one gets the standard string model
with a fluctuating cosmological constant. The Lagrangian $f(R)$, like
Morse function, governs transitions between manifolds with
different topologies.
\end {abstract}
\newpage
\section{Introduction}

There are several motivations to look for a new string model.
One is an old idea that  gauge fields might be related to strings. In
particular, the Nielsen-Olesen string-like solution of the
Higgs model \cite {NO}, 't Hooft large N expansion of gauge theory
\cite {'tH}, the Wilson strong coupling expansion for lattice gauge
theory  \cite {W} and loop space formulation of gauge theory, see for
example  \cite {PAM}, give indications to such relation.
Still it is an open problem
what could be an explicit relation between gauge theory and string theory
and even more  - the standard string theory   is not appropriate
for describing gauge
fields, because of longitude oscillators outside
the critical dimension, for a recent discussion about
this point see  \cite {PolStr}, \cite {Gr}. String theory
also could be applied to solid state physics  and to the
theory of defects
in solid bodies  \cite {KV}. Considerations of modified strings
could help also in understanding  unique properties of the standard string
and in an investigation of the space of all two dimensional field
theories, see \cite {Wi}.

A string model that could correspond to gauge theory  will
be probably more complicated than the standard one and/or will
have additional fields on the worldsheet.
To be useful a modified string model should be tractable
and be natural from a geometrical viewpoint. It is not so easy to
propose such a theory. It seems one can try different modifications
of the standard string model hoping to extend our understanding
of two dmensional field theories and finally to find a string model
which could be appropriate for describing gauge fields and also
to other problems noted above. What could be additional fields
on the worldsheet? From a geometrical viewpoint the most natural
fields are only a metric and a (linear) connection.

In this paper a modification of the  standard string theory
involving a linear connection on the worldsheet is considered.
Recall the action for  bosonic string
\cite {Br}
\begin {equation} 
                                                          \label {1.1}
S=\frac{1}{2}\int _{M} \partial _{\mu}X \partial _{\nu}X
g^{\mu \nu} \sqrt{g}d^{2} \xi
\end   {equation} 
Here $ M$ is a two dimensional manifold with local coordinates
$\xi =(\xi ^{\mu})$  and with a metric $g_{\mu \nu}(\xi)$, $\mu , \nu =1,2,$
$X=(X^{k}(\xi))$, $k=1,...,D$  are coordinates of the string.

 The Euler-Lagrange equations for the action (\ref {1.1}) are
\begin {equation} 
                                                          \label {1.2}
\Box _{g}X=0
\end   {equation} 
\begin {equation} 
                                                          \label {1.3}
\partial _{\mu}X\partial _{\nu}X-\frac{1}{2}
g_{\mu \nu}\partial _{\alpha}X\partial _{\beta}X g^{\alpha \beta}=0
\end   {equation} 
where $\Box _{g}$  is the Laplace-Beltrami operator.

A modification of the string action (\ref {1.1}) which we consider
in this
paper  consists in adding to (\ref {1.1}) a term with a
Lagrangian $f(R)$ ,
where $R$ is the scalar curvature depending on a linear (affine)
connection $\Gamma _{\mu \nu}^{\sigma}$ and a metric
$g_{\mu \nu}$  , $f$
is an analytic function of real variable. So we have the metric
$g_{\mu \nu}$
and the connection $\Gamma _{\mu \nu}^{\sigma}$ as additional fields
on the string worldsheet.  We modify the \it internal \rm
geometry of string introducing the affine connection as a new
geometrical field on the worldsheet.
We are not going to discuss here a modification of \it external \rm
geometry of  string like in models of rigid strings.

The model of 2d gravity
with the Lagrangian $f(R)$  has been considered recently in  \cite {FFV2},
this Lagrangian in an arbitrary dimension $d\geq 3$  has been considered
in  \cite {FFV}. It was shown that for almost arbitrary
analytic Lagrangian $f(R)$
one always gets Einstein equations (if $d\geq 3$) or the
equation of constant
curvature (if $ d=2$).  We will see in this paper that this
universality, i.e. almost independence of equations of motion on $f(R)$
takes place also for string theory.  Actually the traceless of the
energy-momentum tensor of the matter is the crucial property
for the universality at any dimension.

The presence of the metric and the connection on the world sheet looks
rather natural from a geometrical
point of
view.  We will investigate the corresponding classical equations of
motion
and show that in fact equations (\ref {1.2}) and (\ref {1.3})
are preserved
in the modified theory. The whole effect of adding the Lagrangian $f(R)$
to the action (\ref {1.1})  reduces to an interaction of the metric $g$
with the connection
$\Gamma$ which has a form of constant curvature equation.
The space of classical
solutions of the model consists
from sectors of constant curvature $R=c_{i},\, i=1,2...$.
Constants $c_{i}$ are roots of the equation

$$                   f^{\prime} (R)R-f(R)=0$$

If $f^{\prime}(c_i)=0$ one has a sector with the metric interacting
to the connection. If $f^{\prime}(c_i)\ne0$ one has a sector with
the metric interacting
to a vector field which is a solution of an equation for the connection.
 This means that a bifurcation in the space of
solutions takes place.

What is a deep reason for the universality discussed is unclear to me
at the moment. There are important works which could help to clarify
different aspects of universality. Universality of quantum gauge theory
was suggested by Bennet and Nielsen in their works on random dynamics
and gauge glasses \cite {BeNi}. Recently Fairlie, Govaerts and Morosov
\cite {FGM} have found a universal field equation which is invariant under
field redefinition and can be derived from infinity of inequivalent
lagrangians. We see that Einstein equations also enjoy this property
\cite {FFV}. Universality in theory of phase transitions is well
known, for an appropriate consideration in field theory see \cite
{Ar}.

In this paper we consider only classical equations of motion.
Quantization of the model we hope to consider in another paper.
Here we discuss quantization of the model very briefly and  find
that in a semiclassical approximation the model is reduced
to the
standard bosonic string model with a fluctuating cosmological constant.
We will see that the Lagrangian $f(R)$ selects an appropriate
topology of the manifold $M$ and for a good understanding of quantization
of the model one needs a better understanding of topology change in quantum
gravity, for a recent discussion see \cite {Gib,Hor}. In fact the
Lagrangian $f(R)$ plays the role of the Morse function \cite {Mil},
its critical
points are related with change of topology.

Recall that there are other modifications of the action (\ref {1.1}).
In this paper we consider the connection without torsion. But one can add
to (\ref {1.1}) a term describing an interaction of the metric with
a connection with torsion. Adding of this term leads to
a more drastic modification of the standard string theory, in particular,
equation (\ref {1.3}) is modified.The corresponding model of non-critical
strings was considered in \cite {KatVol,Kum}. This model is purely
geometrical one as well as discussed in the present paper. Another
modification of the standard string model which involves an additional
scalar field \cite {Jac} was considered in \cite {Cham}
and has been further generalized in \cite {Licht}.  For a recent
consideration of $R^2$-gravity see \cite {Kaw}.
\section   {The Action and the Equations of Motion}
We consider the following action
\begin {equation} 
                                                          \label {2.1}
S(X,g,\Gamma)=\frac{1}{2}\int _{M} \partial _{\mu}X\partial _{\nu}X
g^{\mu \nu}\sqrt{g} d^{2} \xi
+\int_{M} f(R)\sqrt{g}\ d^2\xi
\end   {equation} 
where $ M $ is a two--dimensional manifold endowed with a metric
$g_{\mu\nu}$ and a symmetric linear connection $\Gamma_{\mu\nu}^\sigma$,
 $\mu ,\nu ,\sigma =1,2$.
One defines the Riemann curvature tensor by
\begin {equation} 
                                                          \label {2.2}
R_{\mu \nu \sigma}^{\lambda}=\partial_{\nu} \Gamma_{\mu
\sigma}^{\lambda} - \partial_{\sigma}\Gamma_{\mu
 \nu}^{\lambda} + \Gamma_{\alpha\nu}^{\lambda} \Gamma_{\mu
\sigma}^{\alpha} - \Gamma_{\alpha \sigma}^{\lambda}
\Gamma_{\mu \nu}^{\alpha}
\end   {equation} 
The Ricci tensor is defined by
\begin {equation} 
                                                          \label {2.3}
R_{\mu \sigma}(\Gamma) = R_{\mu \nu \sigma}^{\nu}
\end   {equation} 
and the scalar curvature
\begin {equation} 
                                                          \label {2.4}
R = R(\Gamma,g)=R_{\mu\sigma}(\Gamma)g^{\mu\sigma}
\end   {equation} 

The function $f$ in  (\ref {2.1}) is a function of one real
variable, which we
assume to be analytic on the real line.

The Euler--Lagrange equations for the action (2.1) with respect to
$X$, $ g$ and $ \Gamma$  can be written in the following form
\begin {equation} 
                                                          \label {2.5}
\Box _{g}X=0,
\end   {equation} 

\begin {equation} 
                                                          \label {2.6}
\frac{1}{2} (\partial _{\mu}X\partial _{\nu}X-\frac{1}{2}
g_{\mu \nu}\partial _{\alpha}X\partial _{\beta}Xg^{\alpha \beta})+
f^{\prime} (R)R_{(\mu \nu )}(\Gamma)-{1\over2}f(R)g_{\mu
\nu}=0
\end   {equation} 

\begin {equation} 
                                                          \label {2.7}
\nabla_{\alpha}(f^{\prime} (R)\sqrt{g} g^{\mu\nu})=0,
\end   {equation} 
where $ \nabla_{\alpha}$ is the covariant derivative with respect
to $ \Gamma$ and $R_{(\mu \nu )}(\Gamma)$ denotes the symmetrical part of
$R_{\mu \nu}(\Gamma)$.

Let us consider equations (\ref {2.6}) and (\ref {2.7}) by using
the same method
as in  \cite {FFV}. Multiplying equation (\ref {2.6}) by $ g^{\mu\nu}$
one obtains
\begin {equation} 
                                                          \label {2.8}
f^{\prime} (R)R-f(R)=0
\end   {equation} 
If equation (\ref {2.8}) is identically satisfied we have
\begin {equation} 
                                                          \label {2.9}
f(R)=cR,
\end   {equation} 
where $c$ is an arbitrary constant.

In all other cases, for a given analytic function $ f(R)$
equation (\ref {2.8})
can have no more than a countable set of solutions
\begin {equation} 
                                                          \label {2.10}
                         R=c_i,
\end   {equation} 
where $c_i$ are constants, $i=1,2,...$.

For a given $c_{i}$ (\ref {2.10}) one can have either
$f^{\prime}(c_i)=0$  or
$f^{\prime}(c_i)\not=0$. First consider
the case $f^{\prime}(c_i)=0$.
Then from (\ref {2.8}) one gets $f(c_i)=0$  and
equation (\ref {2.6})
will take the form (\ref {1.3})
and equation (\ref {2.7}) satisfies identically. Therefore
the full system of equations for $X, g$ and $ \Gamma$ in this case is
\begin {equation} 
                                                          \label {2.11a}
\Box _{g}X=0
\end   {equation} 
\begin {equation} 
                                                          \label {2.11b}
\partial _{\mu}X\partial _{\nu}X-           \frac{1}{2}
g_{\mu \nu}\partial _{\alpha}X\partial _{\beta}Xg^{\alpha \beta}=0
\end   {equation} 
\begin {equation} 
                                                          \label {2.11c}
R(\Gamma, g)=c_{i}.
\end   {equation} 
Equations (\ref {2.11a}) and (\ref {2.11b}) are standard
equation of motion and
constraints for the bosonic string. The metric $g$ interacts
with the connection
$\Gamma$ only by means of equation (\ref {2.11c}).
The whole dependence on the form of the Lagrangian $f$ presents only in the
constant $c_{i}$ which is a root of equation (\ref {2.8}).There are no other
restrictions on $g$ and $\Gamma$.

Now consider the second possibility when $f^{\prime}(c_i)\not=0$.
Then equation (\ref {2.7})  takes the form
\begin {equation} 
                                                          \label {2.12}
\nabla_{\mu}(\sqrt{g} g^{\alpha \beta})=0
\end   {equation} 
Equation (\ref {2.12}) was investigated in  \cite {FFV2} and we present here
the result. Write equation (\ref {2.12}) in the form
\begin {equation} 
                                                          \label {2.13}
\nabla_\mu g_{\alpha\beta}=(-\Gamma ^\sigma_{\mu\sigma}+
{1\over 2}g^{\sigma\tau}\partial_\mu g_{\sigma\tau})g_{\alpha\beta}
\end   {equation} 
Denoting
\begin {equation} 
                                                          \label {2.14a}
B_\mu=\Gamma ^\sigma_{\mu\sigma}-{1\over 2}g^{\sigma\tau}\partial_\mu
 g_{\sigma\tau}
\end   {equation} 
one can rewrite equation (\ref {2.13}) as
\begin {equation} 
                                                          \label {2.14b}
(\nabla_\mu +B_\mu)g_{\alpha\beta}=0
\end   {equation} 
Clearly, the system of equations (\ref {2.14a}) and (\ref {2.14b})
for $\Gamma,g,B$ is equivalent to
equation (\ref {2.12}) or (\ref {2.13}) for $\Gamma$ and $g$.

Let us then consider the system (\ref {2.14a}), (\ref {2.14b}) .
The
general solution of equation (\ref {2.14b}) for the connection
$\Gamma$ is
\begin {equation} 
                                                          \label {2.15}
\Gamma_{\mu \nu}^\sigma
={1\over2}g^{\sigma \alpha}((\partial_{\mu}+B_\mu )g_{\nu \alpha}
+(\partial_{\nu}+B_\nu )g_{\mu \alpha}-(\partial_{\alpha}+B_\alpha )
g_{\mu \nu})
\end   {equation} 
Equation (\ref {2.14a})   follows from (\ref {2.15}).
Let us stress that  we consider
the fields $\Gamma,g$ and $B$ in equations (\ref {2.14a}) and (\ref {2.14b})
as a priori arbitrary fields being the subject to these equations.
Therefore we obtain that the general solution of equation (\ref {2.12})
has the
form (\ref {2.15}) where $g$ and $B$   are arbitrary fields.

Let us remark that equation (\ref {2.12}) is invariant under conformal
(Weyl)
transformation
\begin {equation} 
                                                          \label {2.16}
g_{\alpha \beta} \to e^{\lambda}g_{\alpha \beta}, ~~\Gamma \to \Gamma
\end   {equation} 
and the solution (\ref {2.15}) is invariant under transformations
\begin {equation} 
                                                          \label {2.17}
g_{\alpha \beta} \to e^{\lambda}g_{\alpha \beta}, ~~B_{\mu} \to B_{\mu}-
\partial _{\mu}\lambda.
\end   {equation} 
We need an expression for the Ricci tensor. Using the definition (\ref {2.3})
of the Ricci tensor for the connection
(\ref {2.15}) one gets
\begin {equation} 
                                                          \label {2.18}
R_{\alpha\beta}=R_{(\alpha\beta )}+F_{\alpha\beta}
\end   {equation} 
where
\begin {equation} 
                                                          \label {2.19}
R_{(\alpha\beta)}={1\over2}(R(g)-D_\sigma B^\sigma)g_{\alpha\beta}
\end   {equation} 
and
\begin {equation} 
                                                          \label {2.20}
F_{\alpha\beta}={1\over2}(\partial_\alpha B_\beta - \partial_\beta B_\alpha)
\end   {equation} 
Here $R(g)$ denotes the Ricci  tensor of the metric $g$ and $D_\sigma $ is
the covariant derivative corresponding to the metric $g$. From (\ref {2.8})
(\ref {2.10})  and (\ref {2.19}) one has
\begin {equation} 
                                                          \label {2.21}
R=R_{\alpha \beta}g^{\alpha \beta}=R(g)- D_\sigma  B^\sigma =c_{i}.
\end   {equation} 
Therefore
\begin {equation} 
                                                          \label {2.22}
R_{(\alpha \beta )}= \frac{1}{2}c_{i}g_{\alpha \beta}.
\end   {equation} 
Now by using (\ref {2.8}), (\ref {2.10}) and (\ref {2.22}) we have
\begin {equation} 
                                                          \label {2.23}
f^{\prime} (R)R_{(\mu \nu )}(\Gamma)-{1\over2}f(R)g_{\mu
\nu} =f^{\prime}(c_i)\frac{1}{2} c_{i}g_{\mu \nu }-
{1\over2}f(c_{i})g_{\mu\nu}=0
\end   {equation} 
and equation (\ref {2.6}) reduces to  (\ref {1.3}).

Therefore finally in the case $f^{\prime}(c_{i})\neq 0$ we obtain the
following
system of equations for fields $X$ , $g$ and $B$
\begin {equation} 
                                                          \label {2.24a}
\Box _{g}X=0,
\end   {equation} 
\begin {equation} 
                                                          \label {2.24b}
\partial _{\mu}X\partial _{\nu}X-             {1\over2}
g_{\mu \nu}\partial _{\alpha}X\partial _{\beta}Xg^{\alpha \beta}=0
\end   {equation} 
\begin {equation} 
                                                          \label {2.24c}
R(g)- D_\sigma  B^\sigma =c_{i}.
\end   {equation} 
The connection $\Gamma$ is expressed in terms of $g$ and $B$ by means of
formula (\ref {2.15}).

We find that locally the space of solutions of equations
(\ref {2.5})--(\ref {2.7}) for any analytical function $f$ decomposes into
sectors $H_n$ describing by equations (\ref {2.11a})--(\ref {2.11c})
and  sectors $H_{\varrho}$ describing by equations
(\ref {2.24a})--(\ref {2.24c}). Every sector corresponds to a connected
manifold with a given constant $c_i$ because the curvature is a
continuous function. To take into account all sectors one should
consider the original manifold $M$ in (4) as a disjoint union of
connected manifolds perhaps with different topologies. Therefore
a natural setting for consideration of the action (4) is a collection
of connected manifolds. At classical level there is no interaction
between these manifolds. But after quantization transitions
between sectors with different topologies could appear. One can
compare the Lagrangian $f(R)$ with Morse function \cite {Mil}
describing transitions between manifolds with different topologies.

The case of quadratic function $f(R)$ was considered in detail
in \cite {FFV}. Another simple case admitting an explicit solution
is $f(R)=aR^n+bR+c$.

\section {Conclusion}

We have no intention to consider in this paper the quantization of
the model. Here we make only some preliminary remarks about quantum
properties of the model.The partitition function for the action
(\ref {2.1})
is

\begin {equation} 
                                                          \label {3.1}
Z=\int \exp\{iS(X,g,\Gamma)\}DXDgD\Gamma
\end   {equation} 

An effective action for string variables X after integrating out the
metric and the connection is defined as

\begin {equation} 
                                                        \label {3.2}
exp\{iS_{eff}(X)\}=\int \exp\{iS(X,g,\Gamma)\}DgD\Gamma
\end   {equation} 

In a naive quasiclassical approximation one has a sum over all
critical points of the action, i.e. over all classical solutions
of eqs.(\ref {2.11c}) and (\ref {2.24c}). The space of classical solutions
of these equations
consists from sectors $H_n$ and $ H_{\rho}$ which are
infinite dimensional functional spaces. Therefore in the quasiclassical
approximation still one has functional integrals because there is
sum over the space of all solutions:

\begin {equation} 
                                                         \label {3.3}
e^{iS_{eff}(X)}=\sum_{n} \int e^{iS_{cl}^n} \delta(R(g,
\Gamma)-c_n)DgD\Gamma
+\end   {equation} 
$$\sum_{\varrho} \int e^{iS^{\varrho}_{cl}} \delta(R(g)-
D_{\sigma}B^{\sigma}-c_{\varrho})DgDB,$$
where
$$
S_{cl}^{i}=\frac{1}{2}\int   \partial _{\mu}X\partial _{\nu}X
g^{\mu \nu}\sqrt{g}~ d^{2} \xi +f(c_i)\int\sqrt{g}~ d^{2} \xi
$$

The term $f(c_i)$ is a cosmological constant. So one can
interpret the sums as a string theory with the fluctuating cosmological
constant. Recall that $f(c_n)=0$ in the sector $ H_n$. Note that
this
effect is different from the worm-holes approach \cite {Ha,Co}
where it comes from a summation over nontrivial topologies.

In string theory one considers the sum over different genus
to get a unitary S-matrix in the target space. There are no obvious
reasons why one should have topology change for the Lagrangian
(1) from the point of view of 2-dim gravity. However
if one has the Lagrangian $f(R)$ then even from a point of view
of 2-dim gravity one should take into account all possible
critical points of the action, i.e. $R=c_i, i=1,2...$
like in the Goldstone-Higgs model one should take into account
all critical points of potential energy $V(\phi)$. Therefore
the original manifold $M$ should be a disconnected sum
of connected manifolds with constant curvatures $R=c_i$
and consequently of different topology (genus). The same
arguments one can apply to the Lagrangian $\phi f(R)$
\cite {Licht}  containing an additional scalar
field $\phi$ and to any other Lagrangian leading
to the equations $R=c_i$.

To develope a nonperturbative quantum approach
to this model one needs a lattice formulation
of gravity with a connection. Such a formulation was
suggested in \cite{Cas} and would be very interesting
to use it for this model and also for the string
with torsion \cite {KatVol,Kum}.

\section {Acknowledgments}

I am grateful to I.Aref'eva, M.Caselle, A.D'Adda, D.Fairlie, F.Gliozzi
and M.Katanaev  for stimulating discussions and remarks.

\end{document}